\documentclass[11pt]{article}

\usepackage{amsfonts}
\usepackage{amssymb}
\usepackage{amsmath}
\usepackage{graphicx}

\newcommand{\ket}[1]{|#1\rangle}

\newcommand{\bracket}[2]{\langle #1|#2\rangle}

\def\half{\textstyle{1 \over 2}}
\def\01{\{0,1\}}
\def\x{\times}
\def\e{\varepsilon}

\def\a{\alpha}
\def\b{\beta}

\def\th{^{\mbox{\scriptsize th}}}

\def\ket#1{\mbox{$| #1 \rangle$}}
\def\ts#1{{\textstyle{#1}}}

\def\ni{\noindent}

\def\ee{\vspace*{2mm}}
\def\eee{\vspace*{3mm}}
\def\loud#1{\noindent{\bf #1 }}
\def\newpsi{\psi'}

\newcommand{\op}[1]{\operatorname{#1}}
\newenvironment{mylist}[1]
	{\begin{list}{}{\setlength{\leftmargin}{#1}
	\setlength{\rightmargin}{0.0cm}\setlength{\labelsep}{1.3mm}
	\setlength{\labelwidth}{0.8cm}\setlength{\itemsep}{0.2cm}}}
	{\end{list}}

\newcommand{\qed}{\hfill\rule{2mm}{3mm}}

\newtheorem{theorem}{Theorem}
\newtheorem{lemma}[theorem]{Lemma}

\newtheorem{definition}{Definition}

\begin{document}

\title{\bf\Large Fast parallel circuits for the quantum Fourier transform}

\author{
Richard Cleve\,\thanks{Email: {\tt cleve@cpsc.ucalgary.ca}}
\ \ \ \ \ 
John Watrous\,\thanks{Email: {\tt jwatrous@cpsc.ucalgary.ca}} 
\\ {\small\sl University of Calgary\,}%
\thanks{Department of Computer Science, University of Calgary, Calgary, 
Alberta, Canada T2N 1N4.
Research partially supported by Canada's NSERC.}
}

\date{June 1, 2000}

\maketitle

\begin{abstract}
We give new bounds on the circuit complexity of the quantum Fourier 
transform (QFT).
We give an upper bound of $O(\log n + \log\log (1/\e))$ on the circuit depth 
for computing an approximation of the QFT with respect to the modulus
$2^n$ with error bounded by $\e$.
Thus, even for exponentially small error, our circuits have depth $O(\log n)$.
The best previous depth bound was $O(n)$, even for approximations with
constant error.
Moreover, our circuits have size $O(n \log (n/\e))$.
We also give an upper bound of $O(n (\log n)^2 \log \log n)$ on the 
circuit size of the exact QFT modulo $2^n$, for which the best previous 
bound was $O(n^2)$.

As an application of the above depth bound, we show that Shor's factoring 
algorithm may be based on quantum circuits with depth only $O(\log n)$ 
and polynomial-size, in combination with classical polynomial-time pre- and 
post-processing.
In the language of computational complexity, this implies that factoring 
is in the complexity class $\mbox{ZPP}^{\mbox{\tiny BQNC}}$, where BQNC 
is the class of problems computable with bounded-error probability by 
quantum circuits with poly-logarithmic depth and polynomial size.

Finally, we prove an $\Omega(\log n)$ lower bound on the depth complexity of
approximations of the QFT with constant error.
This implies that the above upper bound is asymptotically optimal (for a 
reasonable range of values of $\e$).

\end{abstract}


\section{Introduction and summary of results}

In this paper we consider the quantum circuit complexity of the {\em quantum 
Fourier transform (QFT)}.
The quantum Fourier transform is the key quantum operation at the heart of 
Shor's quantum algorithms for factoring and computing discrete logarithms 
\cite{Shor97} and the known extensions and variants of these algorithms 
(see, e.g., Kitaev \cite{Kitaev95}, Boneh and Lipton \cite{BonehL95},
Grigoriev \cite{Grigoriev96}, and Cleve, Ekert, Macchiavello, and Mosca 
\cite{CleveE+98}).
The quantum Fourier transform also plays a key role in extensions 
of Grover's quantum searching technique \cite{Grover96}, due to 
Brassard, H{\o}yer, and Tapp \cite{BrassardH+98} and Mosca \cite{Mosca98}.

In order to discuss the quantum Fourier transform in greater detail we recall
the {\em discrete Fourier transform (DFT)}; for a given dimension $m$ the
discrete Fourier transform is a linear operator on $\mathbb{C}^m$ mapping
$(a_0,a_1,\ldots,a_{m-1})$ to $(b_0, b_1, \ldots, b_{m-1})$, where 
\begin{equation}\label{fft}
b_x = \sum_{y = 0}^{m-1} (e^{2 \pi i / m})^{\,x \cdot y} \,a_y.
\end{equation}
The discrete Fourier transform has many important applications in
classical computing, essentially due to the efficiency of the {\em fast Fourier
transform (FFT)\/}, which is an algorithm that computes the DFT with
$O(m \log m)$ arithmetic operations, as opposed to the obvious $O(m^2)$ method.
The FFT algorithm was proposed by Cooley and Tukey in 1965 \cite{CooleyT65},
though its origins can be traced back to Gauss in 1866 \cite{Gauss1866}.
The FFT plays an important role in digital signal processing, and it has been
suggested \cite{GathenG99} as a contender for the second most important
nontrivial algorithm in practice, after fast sorting.
The DFT (and the FFT algorithm) generalize to certain algebraic structures,
such as rings containing primitive $m^{\mbox{\scriptsize th}}$ roots of unity
(which can play the role of $e^{2 \pi i / m}$ in Eq.~\ref{fft}).
This more abstract type of FFT is a principal component in Sch\"{o}nhage 
and Strassen's fast multiplication algorithm \cite{SchonhageS71}, which can be
expressed as circuits of size $O(n \log n \log\log n)$ for multiplying
$n$-bit integers.
For more applications---of which there are many---and historical 
information, see \cite{MaslenR95,Cooley87,HeidemanJ+84}.

The {\em quantum Fourier transform (QFT)} is a unitary operation that
essentially performs the DFT on the amplitude vector of a quantum
state---the QFT maps the quantum state $\sum_{x=0}^{m-1} \a_x \ket{x}$ 
to the state $\sum_{x=0}^{m-1} \b_x \ket{x}$, where 
\begin{equation}
\b_x = \ts{1 \over \sqrt{m}} \sum_{y=0}^{m-1} 
(e^{2 \pi i/ m})^{\,x \cdot y}\,\a_y.
\end{equation}
For certain values of $m$ there are very efficient quantum algorithms for 
the QFT.
The fact that the quantum circuit size can be polynomial in $\log m$ for 
some values of $m$ was first observed by Shor \cite{Shor94} and is of critical 
importance in his polynomial-time algorithms for prime factorization and 
discrete logarithms.
Shor's original method may be described as a ``mixed-radix'' method, and 
is discussed further in Section~\ref{sec:mixed-radix}.
In the particular case where $m=2^n$, there exist quantum circuits performing 
the quantum Fourier transform with $O(n^2)$ gates, which was proved by 
Coppersmith \cite{Coppersmith94} (see also \cite{Cleve94}).
These circuits are based on a recursive description of the QFT that is 
analogous to the description of the DFT exploited by the FFT.
While in some sense these quantum circuits are exponentially faster than the 
classical FFT, the task that they perform is quite different.
The QFT does not explicitly produce any of the values
$\b_0,\b_1,\ldots,\b_{m-1}$ 
as output (nor does it explicitly obtain any of the values 
$\a_0,\a_1,\ldots,\a_{m-1}$ as input).
Intuitively, the difference between performing a DFT and a QFT can be thought
of as being analogous to the difference between computing all the
probabilities that comprise a probability distribution and sampling a
probability distribution---the latter task being frequently much easier.

Coppersmith \cite{Coppersmith94} also proposed quantum circuits that 
approximate the QFT with error bounded by $\e$, and showed that such 
approximations can be computed by circuits of size $O(n \log(n/\e))$ for
modulus $2^n$.
Such approximations can be thought of as unitary operations whose distance
from the QFT (in the operator norm induced by Euclidean distance) is bounded
by $\e$.
Kitaev \cite{Kitaev95} showed how the QFT for an arbitrary modulus $m$ can be
approximated by circuits with size polynomial in $\log (m/\varepsilon)$.
For most information processing purposes, it suffices to use such
approximations of quantum operations (for $\e$ ranging from constant down 
to $1/n^{O(1)}$).
Indeed, since it seems rather implausible to physically implement quantum
gates with perfect accuracy, the need to ultimately consider approximations is
likely inevitable.
Thus, we believe that the most relevant consideration is to approximately 
compute the QFT, though exact computations of the QFT are still of interest 
as part of the mathematical theory of quantum computation.

Moore and Nilsson \cite{MooreN98} showed how to obtain logarithmic-depth 
circuits that perform encoding and decoding for standard quantum 
error-correcting codes.
For the QFT, in both the exact and approximate case, the gates in 
Coppersmith's circuits can be arranged so as to have depth $2n-1$, 
as noted in \cite{MooreN98}, but not less depth than this.
Similarly, the techniques of Shor and of Kitaev have polynomial depth.
Our first result shows that it is possible to compute good approximations 
of the QFT with logarithmic-depth quantum circuits.

\begin{theorem}\sl
\label{theorem:parallel}
For any $n$ and $\varepsilon$ there is a quantum circuit approximating the QFT
modulo $2^n$ with precision $\varepsilon$ that has size $O(n \log(n/\e))$ and
depth $O(\log n + \log\log (1/\varepsilon))$.
\end{theorem}

\ni
By an approximation of a unitary operation $U$ with {\em precision $\e$}, 
we mean a unitary operation $V$ (possibly acting on additional ancilla 
qubits) with the following property.
For any input (pure) quantum state, the Euclidean distance between applying 
$U$ to the state and $V$ to the state is at most $\e$ (in the Hilbert space 
that includes the input/output qubits and the ancilla qubits).
Also, whenever we refer to {\em circuits}, unless otherwise stated, there is 
an implicit assumption that the circuits belong to a logarithmic-space
uniformly generated family in the usual way (via a {\em classical\/} Turing
machine).
In Section~\ref{sec:mixed-radix}, we consider a different approach for
parallelizing Shor's QFT method, which gives somewhat worse bounds.

The proof of Theorem~1 follows the general approach introduced by Kitaev 
\cite{Kitaev95}, with several efficiency improvements as well as 
parallelizations.
In particular, we introduce a new parallel method for performing
multiprecision phase estimation.

We also show that, if size rather than depth is the primary consideration, it
is possible to compute the QFT {\em exactly\/} with a near-linear number of
gates.

\begin{theorem}\sl
\label{theorem:exact}
For any $n$ there is a quantum circuit that exactly computes the QFT modulo
$2^n$ that has size $O(n (\log n)^2 \log\log n)$ and depth $O(n)$.
\end{theorem}

\ni
Theorem~2 is based on a nonstandard recursive description of the QFT
\cite{Cleve94} combined with an asymptotically fast multiplication algorithm
\cite{SchonhageS71}.

There are several reasons why we believe results regarding quantum circuit 
complexity, such as in the above theorems, are important.
First, circuit depth is likely to be particularly relevant in the quantum 
setting for physical reasons.
Perhaps most notably, fault tolerant quantum computation necessarily 
requires parallelization anyway \cite{AharonovB96}---under various noise
models, error correction must continually be applied in parallel to the qubits
of a quantum computer, even when the qubits are doing nothing.
In such models, parallelization saves not only the total amount of time, 
but also the total amount of work.
Furthermore, informally speaking, the depth of a quantum circuit corresponds 
to the amount of time coherence must be preserved, so in addition to saving 
work, parallelization may allow for larger quantum circuits to be implemented 
within systems having shorter decoherence times or using less extensive error 
correction.
A final reason is that such results suggest alternate methods for performing 
various operations, which may in turn suggest or shed light on quantum 
algorithms for other problems or more general methods for improving efficiency 
of quantum algorithms.

It has long been known that the main bottleneck of the quantum portion of
Shor's factoring algorithm is not the QFT, but rather is the modular
exponentiation step.
If it were possible to perform modular exponentiation by classical circuits 
with poly-logarithmic depth and polynomial size then it would be possible to 
implement Shor's factoring algorithm in poly-logarithmic depth with a
polynomial number of qubits.
Although no such algorithm is known for modular exponentiation, we can prove 
the following weaker result, which nevertheless implies that quantum computers 
need only run for poly-logarithmic time for factoring to be feasible.

\begin{theorem}\sl
\label{theorem:factoring}
There is an algorithm for factoring $n$-bit integers that consists of: 
a classical pre-processing stage, computed by a polynomial-size classical 
circuit; 
followed by a quantum information processing stage, computed by an 
$O(\log n)$-depth $O(n^5 (\log n)^2)$-size quantum circuit%
\footnote{In this case, the underlying circuit family is 
{\em polynomial-time\/} uniform rather than logarithmic-space uniform.}; 
followed by a classical post-processing stage, computed by a 
polynomial-size classical circuit.
Furthermore, the size of the quantum circuit can be reduced if a larger 
depth is allowed.
In particular, the size can be reduced to $O(n^3)$ if the depth is increased 
to $O((\log n)^2)$.
\end{theorem}

\ni
If we define the complexity class BQNC as all computational problems that 
can be solved by quantum circuits with poly-logarithmic depth and polynomial 
size---a reasonably natural extension of previous notation (see, e.g., 
\cite{Cook85,MooreN98})---then Theorem~\ref{theorem:factoring} implies that 
the factoring problem is in $\mbox{ZPP}^{\mbox{\tiny BQNC}}$.

Finally, we consider the minimum depth required for approximating the QFT.
It is fairly easy to show that computing the QFT {\em exactly\/} requires 
depth at least $\log n$.
However, this is less clear in the case of approximations---and we exhibit 
a problem related to the QFT whose depth complexity decreases from $\log n$ 
in the exact case to $O(\log\log n)$ for approximations with precision $\e$, 
whenever $\e \in 1/n^{O(1)}$.
Nevertheless, we show the following.

\begin{theorem}\sl
\label{theorem:lowerbounds}
Any quantum circuit consisting of one- and two-qubit gates that approximates 
the QFT with precision $1 \over 10$ or smaller must have depth at least 
$\log n$.
\end{theorem}

\ni
This implies that the depth upper bound in Theorem~\ref{theorem:parallel} 
is asymptotically optimal for a reasonable range of values of $\e$.

The remainder of this paper is organized as follows.
In Section~\ref{sec:notation}, we review some definitions and introduce 
notation that is used in subsequent sections.
In Section~\ref{sec:depth_bound} we prove the depth and size bounds for
quantum circuits approximating the quantum Fourier transform for any
power-of-2 modulus as claimed in Theorem~\ref{theorem:parallel}, and
in Section~\ref{sec:size_bound} we prove the size bound claimed in
Theorem~\ref{theorem:exact} for exactly computing the quantum Fourier
transform.
In Section~\ref{sec:factoring} we prove Theorem~\ref{theorem:factoring}
by demonstrating how Shor's factoring algorithm can be arranged so as to
require only logarithmic-depth quantum circuits.
In Section~\ref{sec:lowerbounds} we prove the lower bound for the QFT in 
Theorem~\ref{theorem:lowerbounds}.
In Section~\ref{sec:other_moduli} we discuss the situation when the modulus
for the quantum Fourier transform is not necessarily a power of 2, including
arbitrary moduli and the special case of ``smooth'' moduli considered
in Shor's original method for performing quantum Fourier transform.
We conclude with Section~\ref{sec:conclusion}, which mentions some
directions for future work relating to this paper.


\section{Definitions and notation}
\label{sec:notation}

{\bf Notation for special quantum states:}
For an $n$-bit modulus $m$, we will identify each $x \in \mathbb{Z}_m$ 
with its binary representation $x_{n-1} \ldots\, x_1 x_0 \in \01^n$.
For $x \in \mathbb{Z}_m$, the state 
$\ket{x} = \ket{x_{n-1} \ldots\, x_1 x_0}$ 
is called a {\em computational basis state}.

For $x \in \mathbb{Z}_m$, the state 
\begin{equation}
\ket{\psi_x} = \ts{1 \over \sqrt{m}} \sum_{y=0}^{m-1} 
(e^{2 \pi i / m})^{x \cdot y} \ket{y},
\end{equation}
is a {\em Fourier basis state\/} with {\em phase parameter $x$}.
As noted in \cite{CleveE+98}, when $m = 2^n$, $\ket{\psi_x}$ can be factored 
as follows
\begin{equation}\label{cemm}
\ket{\psi_{x_{n-1} \ldots\, x_1 x_0}} = \ts{1 \over \sqrt{2^n}} 
(\ket{0} + e^{2 \pi i (0.x_0)}\ket{1})
(\ket{0} + e^{2 \pi i (0.x_1 x_0)}\ket{1})
\cdots
(\ket{0} + e^{2 \pi i (0.x_{n-1} \ldots\, x_1 x_0)}\ket{1}).
\end{equation}
For convenience, we define the state 
\begin{equation}
\ket{\mu_\theta} = 
\ts{\frac{1}{\sqrt{2}}}(\ket{0} + e^{2\pi i\theta}\ket{1}),
\end{equation}
where $\theta$ is a real parameter.
Using this notation, we can rewrite Eq.~\ref{cemm} as 
\begin{equation}
\label{cemm2}
\ket{\psi_{x_{n-1} \ldots\, x_1 x_0}} = 
\ket{\mu_{0.x_0}}\ket{\mu_{0.x_1 x_0}} \cdots 
\ket{\mu_{0.x_{n-1} \ldots\, x_1 x_0}}.
\end{equation}

\ni
{\bf Definition of the QFT:}
The {\em quantum Fourier transform (QFT)\/} is the unitary 
operation that maps $\ket{x}$ to $\ket{\psi_x}$ (for all 
$x \in \mathbb{Z}_m$).\ee

\ni
{\bf Mappings related to the QFT:}
A {\em quantum Fourier state computation (QFS)\/} is any 
unitary operation that maps $\ket{x}\ket{0}$ to 
$\ket{x}\ket{\psi_x}$ (for all $x \in \mathbb{Z}_m$).
When the input is a computational basis state, this computes the 
corresponding Fourier state, but without erasing the input.
We refer to approximations of a QFS as {\em Fourier state 
estimation}.
A {\em quantum Fourier phase computation (QFP)\/} is any 
unitary operation that maps $\ket{\psi_x}\ket{0}$ to 
$\ket{\psi_x}\ket{x}$ (for all $x \in \mathbb{Z}_m$).
When the input is a Fourier basis state, this computes the 
corresponding phase parameter, but without erasing the input.
We refer to approximations of a QFP as {\em Fourier phase 
estimation}.
As pointed out by Kitaev \cite{Kitaev95}, the QFT can be computed 
by composing a QFS and the inverse of a QFP: 
$\ket{x}\ket{0} \mapsto \ket{x}\ket{\psi_x} \mapsto 
\ket{0}\ket{\psi_x}$.\ee

\ni
{\bf Quantum gates:}
All of the quantum circuits that we construct will be composed of 
three types of unitary gates.
One is the one-qubit {\em Hadamard\/} gate, $H$, which maps 
$\ket{x}$ to ${1 \over \sqrt{2}}(\ket{0} + (-1)^x \ket{1})$ 
(for $x \in \01$).
Another is the one-qubit {\em phase shift\/} gate, $P(\theta)$, 
where $\theta$ is a parameter of the form $x / 2^n$ 
(for $x \in \mathbb{Z}_{2^n}$).
$P(\theta)$ maps $\ket{x}$ to $e^{2 \pi i\theta x}\ket{x}$ 
(for $x \in \01$).
Finally, we use two-qubit {\em controlled}-phase shift gates, 
$\mbox{controlled-{\it P\/}}(\theta)$ ($\mbox{c-{\it P\/}}(\theta)$ for
short), which map $\ket{x}\ket{y}$ to $e^{2 \pi i\theta x y}\ket{x}\ket{y}$ 
(for $x, y \in \01$).
Note that this set is universal, and in particular that any (classical)
reversible circuit can be composed of these gates.


\section{New depth bounds for the QFT}
\label{sec:depth_bound}

The main purpose of this section is to prove Theorem~\ref{theorem:parallel}.

First, we review the approach of Kitaev \cite{Kitaev95} for performing the 
QFT for an arbitrary modulus $m$.
By linearity, it is sufficient to give a circuit that operates correctly 
on computational basis states.
Given a computational basis state $\ket{x}$, first create the Fourier 
basis state with phase parameter $x$ (which can be done easily if $\ket{x}$ 
is not erased in the process).
The system is now in the state $\ket{x}\ket{\psi_x}$.
Now, by performing Fourier phase estimation, the state $\ket{x}\ket{\psi_x}$ 
can be approximated from the state $\ket{0}\ket{\psi_x}$.
Therefore, by performing the inverse of Fourier phase estimation on 
the state $\ket{x}\ket{\psi_x}$, a good estimate of the state 
$\ket{0}\ket{\psi_x}$ is obtained.

The particular phase estimation procedure used by Kitaev does not readily
parallelize, but, in the case where the modulus is a power of 2, we give a
new phase estimation procedure that does parallelize.
This procedure requires several copies of the Fourier basis state rather 
than just one.
To insure that the entire process parallelizes, we must parallelize the 
creation of the Fourier basis state as well as the process of copying and 
uncopying this state.

The basic steps of our technique are as follows: 
\begin{enumerate}
\item
Creation of the Fourier basis state, which is the mapping
\[
\ket{x}\ket{0} \mapsto \ket{x}\ket{\psi_x}.
\]
\item
Copying the Fourier basis state, which is the mapping
\[
\ket{\psi_x}\ket{0} \cdots \ket{0} \mapsto
\ket{\psi_x}\ket{\psi_x} \cdots \ket{\psi_x}.
\]
\item
Erasing the computational basis state by means of estimating the phase of the
Fourier basis state, which is the mapping 
\[
\ket{x}\ket{\psi_x}\ket{\psi_x} \cdots \ket{\psi_x} \mapsto 
\ket{0}\ket{\psi_x}\ket{\psi_x} \cdots \ket{\psi_x}.
\]
\item 
Reverse step 2, which is the mapping
\[
\ket{\psi_x}\ket{\psi_x} \cdots \ket{\psi_x}\mapsto \ket{\psi_x}\ket{0}
\cdots \ket{0}.
\]
\end{enumerate}

\noindent
Each of these components is discussed in detail in the subsections that follow.
Throughout we assume the modulus is $m = 2^n$.


\subsection{Parallel Fourier state computation and estimation}
\label{subsection:create}

The first step is the creation of the Fourier basis state corresponding to a
given computational basis state $\ket{x}$.
This corresponds to the mapping
\begin{equation}
\ket{x}\ket{0} \mapsto \ket{x}\ket{\psi_x}.
\label{eq:create}
\end{equation}

First let us consider a circuit that performs this transformation 
exactly.
By Eq.~\ref{cemm2} (equivalently, Eq.~\ref{cemm}), it suffices to compute 
the states 
$\ket{\mu_{0.x_0}}, \ket{\mu_{0.x_1 x_0}}, \ldots, 
\ket{\mu_{0.x_{n-1} \ldots\, x_1 x_0}}$ individually.

The circuit suggested by Figure~\ref{fig:exact_prep} performs the required 
transformation for $\ket{\mu_{0.x_j \ldots\, x_0}}$.
\begin{figure}[!ht]
\rule{1cm}{0cm}
\setlength{\unitlength}{2565sp}%
\begingroup\makeatletter\ifx\SetFigFont\undefined
\def\x#1#2#3#4#5#6#7\relax{\def\x{#1#2#3#4#5#6}}%
\expandafter\x\fmtname xxxxxx\relax \def\y{splain}%
\ifx\x\y   
\gdef\SetFigFont#1#2#3{%
  \ifnum #1<17\tiny\else \ifnum #1<20\small\else
  \ifnum #1<24\normalsize\else \ifnum #1<29\large\else
  \ifnum #1<34\Large\else \ifnum #1<41\LARGE\else
     \huge\fi\fi\fi\fi\fi\fi
  \csname #3\endcsname}%
\else
\gdef\SetFigFont#1#2#3{\begingroup
  \count@#1\relax \ifnum 25<\count@\count@25\fi
  \def\x{\endgroup\@setsize\SetFigFont{#2pt}}%
  \expandafter\x
    \csname \romannumeral\the\count@ pt\expandafter\endcsname
    \csname @\romannumeral\the\count@ pt\endcsname
  \csname #3\endcsname}%
\fi
\fi\endgroup
\begin{picture}(8910,5495)(376,-6744)
\thinlines
\put(4801,-3061){\circle*{150}}
\put(4801,-3661){\circle*{150}}
\put(4801,-1561){\circle*{150}}
\put(4801,-2461){\circle*{150}}
\put(4801,-2161){\circle{150}}
\put(4801,-2761){\circle{150}}
\put(4801,-3361){\circle{150}}
\put(4801,-3961){\circle{150}}
\put(4801,-1561){\line( 0,-1){675}}
\put(4801,-2461){\line( 0,-1){375}}
\put(4801,-3061){\line( 0,-1){375}}
\put(4801,-3661){\line( 0,-1){375}}
\put(4201,-1561){\circle*{150}}
\put(4201,-3061){\circle*{150}}
\put(4201,-3661){\circle{150}}
\put(4201,-2461){\circle{150}}
\put(4201,-1561){\line( 0,-1){975}}
\put(4201,-3061){\line( 0,-1){675}}
\put(3601,-1561){\circle*{150}}
\put(3601,-3061){\circle{150}}
\put(3601,-1561){\line( 0,-1){1575}}
\put(7201,-3061){\circle*{150}}
\put(7201,-3661){\circle*{150}}
\put(7201,-1561){\circle*{150}}
\put(7201,-2461){\circle*{150}}
\put(7201,-2161){\circle{150}}
\put(7201,-2761){\circle{150}}
\put(7201,-3361){\circle{150}}
\put(7201,-3961){\circle{150}}
\put(7201,-1561){\line( 0,-1){675}}
\put(7201,-2461){\line( 0,-1){375}}
\put(7201,-3061){\line( 0,-1){375}}
\put(7201,-3661){\line( 0,-1){375}}
\put(7801,-1561){\circle*{150}}
\put(7801,-3061){\circle*{150}}
\put(7801,-3661){\circle{150}}
\put(7801,-2461){\circle{150}}
\put(7801,-1561){\line( 0,-1){975}}
\put(7801,-3061){\line( 0,-1){675}}
\put(8401,-1561){\circle*{150}}
\put(8401,-3061){\circle{150}}
\put(8401,-1561){\line( 0,-1){1575}}
\put(5401,-4561){\circle*{150}}
\put(5551,-4861){\circle*{150}}
\put(5701,-5161){\circle*{150}}
\put(5851,-5461){\circle*{150}}
\put(6001,-5761){\circle*{150}}
\put(6151,-6061){\circle*{150}}
\put(6301,-6361){\circle*{150}}
\put(6451,-6661){\circle*{150}}
\put(5401,-1561){\circle*{150}}
\put(5551,-2161){\circle*{150}}
\put(5701,-2461){\circle*{150}}
\put(5851,-2761){\circle*{150}}
\put(6001,-3061){\circle*{150}}
\put(6151,-3361){\circle*{150}}
\put(6301,-3661){\circle*{150}}
\put(6451,-3961){\circle*{150}}
\put(2401,-1861){\framebox(600,600){\large H}}
\put(1801,-1561){\line( 1, 0){600}}
\put(5401,-1561){\line( 0,-1){3000}}
\put(5551,-2161){\line( 0,-1){2700}}
\put(5701,-2461){\line( 0,-1){2700}}
\put(5851,-2761){\line( 0,-1){2700}}
\put(6001,-3061){\line( 0,-1){2700}}
\put(6151,-3361){\line( 0,-1){2700}}
\put(6301,-3661){\line( 0,-1){2700}}
\put(6451,-3961){\line( 0,-1){2700}}
\put(3001,-1561){\line( 1, 0){6000}}
\put(1801,-2161){\line( 1, 0){7200}}
\put(1801,-2461){\line( 1, 0){7200}}
\put(1801,-2761){\line( 1, 0){7200}}
\put(1801,-3061){\line( 1, 0){7200}}
\put(1801,-3361){\line( 1, 0){7200}}
\put(1801,-3661){\line( 1, 0){7200}}
\put(1801,-3961){\line( 1, 0){7200}}
\put(1801,-4561){\line( 1, 0){7200}}
\put(1801,-4861){\line( 1, 0){7200}}
\put(1801,-5161){\line( 1, 0){7200}}
\put(1801,-5461){\line( 1, 0){7200}}
\put(1801,-5761){\line( 1, 0){7200}}
\put(1801,-6061){\line( 1, 0){7200}}
\put(1801,-6361){\line( 1, 0){7200}}
\put(1801,-6661){\line( 1, 0){7200}}
\put(9500,-1681){\makebox(0,0)[lb]{$\ket{\mu_{0.x_j\cdots x_0}}$}}
\put(9500,-3166){\makebox(0,0)[lb]{$\ket{0^{j}}$}}
\put(9500,-4700){\makebox(0,0)[lb]{$\ket{x_j}$}}
\put(800,-4700){\makebox(0,0)[lb]{$\ket{x_j}$}}
\put(9600,-5700){\makebox(0,0)[lb]{$\vdots$}}
\put(900,-5700){\makebox(0,0)[lb]{$\vdots$}}
\put(800,-6800){\makebox(0,0)[lb]{$\ket{x_0}$}}
\put(9500,-6800){\makebox(0,0)[lb]{$\ket{x_0}$}}
\put(800,-3166){\makebox(0,0)[lb]{$\ket{0^{j}}$}}
\put(800,-1681){\makebox(0,0)[lb]{$\ket{0}$}}
\end{picture}\vspace{2mm}
\caption{Quantum circuit for the exact preparation of 
$\ket{\mu_{0.x_j \cdots\, x_0}}$.}
\label{fig:exact_prep}
\end{figure}
In this figure we have not labelled the controlled phase shift gates, 
$\mbox{c-{\it P\/}}(\theta)$ (such gates are defined in 
Section~\ref{sec:notation}), which are the gates in the center drawn as two 
solid circles connected by a line.
In the above case, the phase $\theta$ depends on $j$ and on the particular 
qubit of $\ket{x_{n-1} \ldots\, x_1 x_0}$ on which the gate acts.
The value of $\theta$ for the controlled phase shift acting on $\ket{x_i}$ 
is $2^{i-j-1}$ (for $i \in \{0,1,\ldots,j\}$).  From this, it may be verified
that the circuit acts as indicated.
The depth of this circuit is $O(\log n)$ and the size is $O(n)$.

If such a circuit is to be applied for each value of $j \in 
\{0,1,\ldots,n-1\}$, in order to perform the mapping (\ref{eq:create}), 
then the qubits $\ket{x_{n-1}},  \ldots\, \ket{x_1}, \ket{x_0}$ must 
first be copied several times ($n-i$ times for $\ket{x_i}$) 
to allow the controlled phase shift gates to operate in parallel.
This may be performed (and inverted appropriately) in size $O(n^2)$ and 
depth $O(\log n)$ in the most obvious way.
We conclude that the transformation (\ref{eq:create}) can be performed by
circuits of size $O(n^2)$ and depth $O(\log n)$ in the exact case.

In order to reduce the size of the circuit in the approximate case, we 
use a similar procedure, except we only perform the controlled phase shifts 
when the phase $\theta$ is significant.
An illustration of such a circuit is given in Figure~\ref{fig:approx_prep}.
\begin{figure}[!ht]
\rule{1cm}{0cm}
\setlength{\unitlength}{2565sp}%
\begingroup\makeatletter\ifx\SetFigFont\undefined
\def\x#1#2#3#4#5#6#7\relax{\def\x{#1#2#3#4#5#6}}%
\expandafter\x\fmtname xxxxxx\relax \def\y{splain}%
\ifx\x\y   
\gdef\SetFigFont#1#2#3{%
  \ifnum #1<17\tiny\else \ifnum #1<20\small\else
  \ifnum #1<24\normalsize\else \ifnum #1<29\large\else
  \ifnum #1<34\Large\else \ifnum #1<41\LARGE\else
     \huge\fi\fi\fi\fi\fi\fi
  \csname #3\endcsname}%
\else
\gdef\SetFigFont#1#2#3{\begingroup
  \count@#1\relax \ifnum 25<\count@\count@25\fi
  \def\x{\endgroup\@setsize\SetFigFont{#2pt}}%
  \expandafter\x
    \csname \romannumeral\the\count@ pt\expandafter\endcsname
    \csname @\romannumeral\the\count@ pt\endcsname
  \csname #3\endcsname}%
\fi
\fi\endgroup
\begin{picture}(8850,4224)(451,-5473)
\thinlines
\put(5401,-1561){\circle*{150}}
\put(5551,-2161){\circle*{150}}
\put(5701,-2461){\circle*{150}}
\put(5851,-2761){\circle*{150}}
\put(4201,-1561){\circle*{150}}
\put(4201,-2461){\circle{150}}
\put(4801,-1561){\circle*{150}}
\put(4801,-2461){\circle*{150}}
\put(4801,-2161){\circle{150}}
\put(4801,-2761){\circle{150}}
\put(7201,-1561){\circle*{150}}
\put(7201,-2461){\circle*{150}}
\put(7201,-2161){\circle{150}}
\put(7201,-2761){\circle{150}}
\put(7801,-1561){\circle*{150}}
\put(7801,-2461){\circle{150}}
\put(5401,-3961){\circle*{150}}
\put(5551,-4261){\circle*{150}}
\put(5701,-4561){\circle*{150}}
\put(5851,-4861){\circle*{150}}
\put(2401,-1861){\framebox(600,600){\large H}}
\put(1801,-1561){\line( 1, 0){600}}
\put(3001,-1561){\line( 1, 0){6000}}
\put(1801,-2161){\line( 1, 0){7200}}
\put(1801,-2461){\line( 1, 0){7200}}
\put(1801,-2761){\line( 1, 0){7200}}
\put(1801,-4561){\line( 1, 0){7200}}
\put(1801,-4861){\line( 1, 0){7200}}
\put(4201,-1561){\line( 0,-1){975}}
\put(4801,-1561){\line( 0,-1){675}}
\put(4801,-2461){\line( 0,-1){375}}
\put(7201,-1561){\line( 0,-1){675}}
\put(7201,-2461){\line( 0,-1){375}}
\put(7801,-1561){\line( 0,-1){975}}
\put(1801,-3961){\line( 1, 0){7200}}
\put(1801,-4261){\line( 1, 0){7200}}
\put(5401,-1561){\line( 0,-1){2400}}
\put(5551,-2161){\line( 0,-1){2100}}
\put(5701,-2461){\line( 0,-1){2100}}
\put(5851,-2761){\line( 0,-1){2100}}
\put(9300,-1700){\makebox(0,0)[lb]{$\ket{\mu_{0.x_j\cdots
	x_{j-k+1}}}$}}
\put(650,-1700){\makebox(0,0)[lb]{$\ket{0}$}}
\put(9300,-5000){\makebox(0,0)[lb]{$\ket{x_{j-k+1}}$}}
\put(9300,-4100){\makebox(0,0)[lb]{$\ket{x_j}$}}
\put(650,-5000){\makebox(0,0)[lb]{$\ket{x_{j-k+1}}$}}
\put(800,-4550){\makebox(0,0)[lb]{$\vdots$}}
\put(9450,-4550){\makebox(0,0)[lb]{$\vdots$}}
\put(650,-4100){\makebox(0,0)[lb]{$\ket{x_j}$}}
\put(650,-2600){\makebox(0,0)[lb]{$\ket{0^{k-1}}$}}
\put(9300,-2600){\makebox(0,0)[lb]{$\ket{0^{k-1}}$}}
\end{picture}\vspace{-2mm}
\caption{Quantum circuit for the approximate preparation of
$\ket{\mu_{0.x_j \ldots\, x_0}}$.}
\label{fig:approx_prep}
\end{figure}
Here $k$ denotes the number of significant phase 
shift gates that are used.
The condition 
$\|\ket{\mu_{0.x_j \cdots\, x_0}}-\ket{\mu_{0.x_j \cdots\, x_{j-k+1}}}\| 
\in (\e/n)^{O(1)}$ occurs when $k \in O(\log (n/\e))$.
With such a setting of $k$, the precision of the approximation of 
$\ket{\mu_{0.x_{n-1} \cdots\, x_0}} \cdots \ket{\mu_{0.x_0}}$ can 
be $O(\e)$.
Note that the size of the resulting circuit is $O(n\log (n/\e))$ and the 
depth is $O(\log\log(n/\e))$.


\subsection{Copying a Fourier state in parallel}
\label{subsec:copy}

In this section, we show how to efficiently produce $k$ copies of 
an $n$-qubit Fourier state from one copy.
This is a unitary operation that acts on $k$ $n$-qubit registers 
(thus $kn$ qubits in all) and maps 
$\ket{\psi_x}\ket{0^n}\cdots\ket{0^n}$ 
to $\ket{\psi_x}\ket{\psi_x}\cdots\ket{\psi_x}$ 
for all $x \in \01^n$.
The copying circuit will be exact and have size $O(kn)$ and 
depth $O(\log(kn))$.
The setting of $k$ will be $O(\log(n/\e))$.

Let us begin by considering the problem of producing two copies 
of a Fourier state from one.
First, define the {\em (reversible) addition\/} and 
{\em (reversible) subtraction\/} operations as the mappings 
\begin{eqnarray*}
\ket{x}\ket{y} & \mapsto & \ket{x}\ket{y+x} \\
\ket{x}\ket{y} & \mapsto & \ket{x}\ket{y-x} 
\end{eqnarray*}
(respectively), where $x, y \in \01^n$ and additions and 
subtractions are performed as integers modulo $2^n$.
By appealing to classical results about the complexity of 
arithmetic \cite{Ofman63}, one can construct quantum circuits 
of size $O(n)$ and depth $O(\log n)$ for these operations 
(using an ancilla of size $O(n)$).

It is straightforward to show that applying a subtraction to 
the state $\ket{\psi_x}\ket{\psi_y}$ results in the state  
$\ket{\psi_{x+y}}\ket{\psi_y}$.
Also, the state $\ket{\psi_0}$ can be obtained from $\ket{0^n}$ 
by applying a Hadamard transform independently to each qubit.
Therefore, the copying operation can begin with a state of the 
form $\ket{0^n}\ket{\psi_x}$ and consist of these two steps:
\begin{enumerate}
\item
Apply $H$ to each of the first $n$ qubits.
\item
Apply the subtraction operation to the $2n$ qubits.
\end{enumerate}
The resulting state will be $\ket{\psi_x}\ket{\psi_x}$.

An obvious method for computing $k$ copies of a Fourier state 
is to repeatedly apply the above doubling operation.
This will result in a quantum circuit of size $O(kn)$; 
however, its depth will be $O((\log k)(\log n))$, which 
is too large for our purposes.

The depth bound can be improved to $O(\log(kn))$ by applying 
other classical circuit constructions to efficiently implement 
the {\em (reversible) prefix addition\/} and 
{\em (reversible) telescoping subtraction\/} operations, 
which are the mappings 
\begin{eqnarray*}
\ket{x_1}\ket{x_2} \cdots \ket{x_k} & \mapsto & 
\ket{x_1}\ket{x_1+x_2} \cdots \ket{x_1+x_2+\cdots+x_k} \\
\ket{x_1}\ket{x_2} \cdots \ket{x_k} & \mapsto & 
\ket{x_1}\ket{x_2-x_1} \cdots \ket{x_k - x_{k-1}}
\end{eqnarray*}
(respectively), where $x_1, x_2, \ldots, x_k \in \01^n$.
Before addressing the issue of efficiently implementing 
these operations, let us note that the copying operation 
can be performed by starting with the state 
$\ket{0^n}\cdots\ket{0^n}\ket{\psi_x}$ and performing these 
two steps: 
\begin{enumerate}
\item
Apply $H$ to all of the first $(k-1)n$ qubits.
\item
Apply the telescoping subtraction operation to the 
$k n$ qubits.
\end{enumerate}
The resulting state will be $\ket{\psi_x}\cdots\ket{\psi_x}$.

Now, to implement the prefix addition and telescoping subtraction, 
note that they are inverses of each other.
This means that it is sufficient to implement each one efficiently 
by a classical (nonreversible) circuit, and then combine these 
to produce a reversible circuit by standard techniques in reversible 
computing \cite{Bennett73}.
The telescoping subtraction clearly consists of $k-1$ subtractions 
that can be performed in parallel, so the nonreversible size and depth 
bounds are $O(k n)$ and $O(\log n)$ respectively.

The prefix addition is a little more complicated.
It relies on a combination of well-known tools in classical circuit 
design.
One of them is the following general result of Ladner and Fischer 
\cite{LadnerF80} about parallel prefix computations.

\begin{theorem}[Ladner and Fischer]\sl
\label{theorem:pp}
For any associative binary operation $\circ$, the mapping 
\begin{equation}
(x_1,\ x_2,\ \ldots,\ x_k) \mapsto 
(x_1,\ x_1 \circ x_2,\,\ldots,\, x_1 \circ x_2 \circ \cdots \circ x_k)
\end{equation}
can be computed by a circuit consisting of 
$(x,\, y) \mapsto (x,\, x \circ y)$ gates that has size 
$O(k)$ and depth $O(\log k)$.
\end{theorem}

Another tool is the so-called {\em three-two adder}, which is 
a circuit that takes three $n$-bit integers $x, y, z$ as input 
and produces two $n$-bit integers $s, c$ as output, such that 
$x+y+z = s+c$ (recall that addition is in modulo $2^n$ 
arithmetic).
It is remarkable that a three-two adder can be implemented with 
{\em constant depth\/} and size $O(n)$.
By combining two three-two adders, one can implement a 
size $O(n)$ and depth $O(1)$ {\em four-two adder}, that 
performs the mapping $(x,y,z,w) \mapsto (x,y,s,c)$, 
where $x+y+z+w = s+c$.
Now, consider the {\em pairwise representation\/} of each $n$-bit 
integer $z$ as a pair of two $n$-bit integers 
$(z^{\prime},z^{\prime\prime})$ such that 
$z = z^{\prime} + z^{\prime\prime}$.
This representation is not unique, but it is easy to convert 
to and from the pairwise representation: the respective 
mappings are $z \mapsto (z,0^n)$ and 
$(z^{\prime},z^{\prime\prime}) 
\mapsto z^{\prime} + z^{\prime\prime}$.
The useful observation is that the four-two adder performs 
integer addition in the pairwise representation scheme, and 
it does so in constant depth and size $O(n)$.

Now, the following procedure computes prefix addition in 
size $O(k n)$ and depth $O(\log k + \log n) = O(\log(k n))$.
The input is $(x_1,\,x_2,\,\ldots,\,x_k)$.
\begin{enumerate}
\item
Convert the $k$ integers into their pairwise representation.
\item
Apply the parallel prefix circuit of Theorem~\ref{theorem:pp}
to perform the prefix additions in the pairwise 
representation scheme.
\item
Convert the $k$ integers from their pairwise representation 
to their standard form.
\end{enumerate}
The output will be 
$(x_1,\ x_1+x_2,\ \ldots,\ x_1 + x_2 + \cdots + x_k)$, 
as required.

Note that step 4 of the main algorithm has a circuit of identical size and
depth to the one just described, as it is simply its inverse.


\subsection{Estimating the phase of a Fourier state in parallel}
\label{subsec:phase_estimation}

Finally, we will discuss the third step of the main algorithm, which
corresponds to the mapping
\begin{equation}
\ket{\psi_x}\ket{\psi_x}\cdots\ket{\psi_x}\ket{x}\mapsto
\ket{\psi_x}\ket{\psi_x}\cdots\ket{\psi_x}\ket{0}
\label{eq:mapping3}
\end{equation}
for $x\in\{0,1\}^n$.
The number of copies of $\ket{\psi_x}$ required for this step depends on the
error bound $\varepsilon$; we will require $k\in O(\log(n/\varepsilon))$
copies.
As discussed in subsection~\ref{subsection:create}, any Fourier basis state
$\ket{\psi_x}$ may be decomposed as
$\ket{\psi_x} = \ket{\mu_{x2^{-1}}}
\ket{\mu_{x2^{-2}}}\cdots
\ket{\mu_{x2^{-n}}}$.
Thus, we may assume that we have $k$ copies of each of the states
$\ket{\mu_{x2^{-j}}}$.

First, for each $j = 1,\ldots,n$, the circuit will simulate measurements of
the $k$ copies of $\ket{\mu_{x2^{-j}}}$ (in the bases discussed below) in
order to obtain an approximation $l_j/4$ to the fractional part of $2^{-j}x$.
The approximation is with respect to the function $|\cdot|_1$ defined as
\[
|y|_1 = \op{min}\left\{z\in[0,1)\,:\,\mbox{either $y-z\in\mathbb{Z}$ or
$y+z\in\mathbb{Z}$ }\right\}
\]
for $y\in\mathbb{R}$ (i.e., ``modulo 1'' distance).
With high probability the approximations will result in $l_1,\ldots,l_n$
satisfying $|l_j/4-2^{-j}x|_1 < \frac{1}{4}$ for each $j$.
The (simulated) measurements can be performed in parallel, and each $l_j$ will
be determined by considering the mode of the outcomes of the measurements
and thus can be computed in parallel as well.
Next the circuit will reconstruct an approximation $\widetilde{x}$
to $x$ (in parallel) from $l_1,\ldots,l_n$.
The circuit then XORs this value of $\widetilde{x}$ to the register containing
$x$, thereby ``erasing'' it with high probability.
Finally, the circuit inverts the computation of this $\widetilde{x}$ to clean
up any garbage from the computation.
As in subsection~\ref{subsec:copy}, standard techniques may be used to
implement these computations as reversible circuits.
We now describe each of the above steps in more detail.

Let us first recall the following fact from probability theory
(see, e.g., Goldreich \cite{Goldreich99}).
If $X_1,\ldots,X_t$ are independent Bernoulli trials with probability
$p_X$ of success and $Y_1,\ldots,Y_t$ are independent Bernoulli trials
with $p_Y$ of success, where $p_X<p_Y$, then
\[
\op{Pr}\left[\sum_{i=1}^t X_i \geq \sum_{i=1}^t Y_i\right]
< 2 e^{-(p_Y-p_X)^2\,t/2}.
\]

Now, define
\[
\begin{array}{ll}
\ket{b_0} = \ts{\frac{1}{\sqrt{2}}}\ket{0} + \ts{\frac{1}{\sqrt{2}}}\ket{1}
= \ket{\mu_0}, &
\ket{b_1} = \ts{\frac{1}{\sqrt{2}}}\ket{0} + \ts{\frac{i}{\sqrt{2}}}\ket{1}
= \ket{\mu_{1 \over 4}},\\[2mm]
\ket{b_2} = \ts{\frac{1}{\sqrt{2}}}\ket{0} - \ts{\frac{1}{\sqrt{2}}}\ket{1}
= \ket{\mu_{1 \over 2}}, &
\ket{b_3} = \ts{\frac{1}{\sqrt{2}}}\ket{0} - \ts{\frac{i}{\sqrt{2}}}\ket{1}
= \ket{\mu_{3 \over 4}},
\end{array}
\]
and consider measurements of the states $\ket{\mu_{x2^{-j}}}$ in the bases
$\{\ket{b_0},\ket{b_2}\}$ and $\{\ket{b_1},\ket{b_3}\}$ (these measurements 
correspond to measurements of the Pauli operators $\sigma_x$ and $\sigma_y$, 
respectively).
In particular, given that we have $k$ copies of each $\ket{\mu_{x2^{-j}}}$, we
suppose that each of the above two measurements is performed independently on
$k/2$ of the copies.
Let $l_j\in\{0,1,2,3\}$ represent the basis state that occurs with the highest
frequency in these measurements for each $j$, breaking ties arbitrarily.
We claim that the inequality $|l_j/4-2^{-j}x|_1<\frac{1}{4}$ is satisfied with
high probability:
\begin{equation}
\op{Pr}\left[\left|l_j/4-2^{-j}x\right|_1 \geq \ts{\frac{1}{4}}\right]
<4e^{-k/8}.
\label{eq:prob_bound}
\end{equation}

To prove that the inequality (\ref{eq:prob_bound}) holds, let us suppose that
$x$ and $j$ are fixed, and let us define
\[
p_0 = |\bracket{b_0}{\mu_{x2^{-j}}}|^2,\;\;\;
p_1 = |\bracket{b_1}{\mu_{x2^{-j}}}|^2,\;\;\;
p_2 = |\bracket{b_2}{\mu_{x2^{-j}}}|^2,\;\;\;
p_3 = |\bracket{b_3}{\mu_{x2^{-j}}}|^2.
\]
These are the probabilities associated with the above measurements, meaning
that the probability that a measurement of $\ket{\mu_{x2^{-j}}}$ in the
$\{\ket{b_0},\ket{b_2}\}$ basis yields 0 is $p_0$, the probability
that the measurement yields 2 is $p_2$, and similar for $p_1$ and $p_3$
when the measurement in the $\{\ket{b_1},\ket{b_3}\}$ basis is performed.
Now, note the following two facts: (i) it must be that
$\op{max}\{p_0,p_1,p_2,p_3\}\geq 1/2 + \sqrt{2}/4$ (for any choice of $x$ and
$j$), and (ii) if $|l/4 - 2^{-j}x|_1\geq \frac{1}{4}$, then we must have
$p_l\leq 1/2$.
Therefore, if the inequality is not satisfied for some $j$ (i.e.,
if $|l_j/4 - 2^{-j}x|_1\geq \frac{1}{4}$), then it must be the case that
$p_{l'} - p_{l_j} \geq \sqrt{2}/4$ for some different value of $l'\not=l_j$.
Based on the inequalities above, we conclude that a very improbable event
has taken place: the probability of the result $l_j$ appearing more frequently
than $l'$ is at most $2 e^{-k/8}$.
Unless $|2^{-j}x|_1\in\{0,\frac{1}{4},\frac{1}{2},\frac{3}{4}\}$ there are at
most two values of $l_j$ that give $|l_j/4 - 2^{-j}x|_1\geq \frac{1}{4}$, 
and so in this case we conclude that (\ref{eq:prob_bound}) holds.
(In the special case $|2^{-j}x|_1\in\{0,\frac{1}{4},\frac{1}{2},\frac{3}{4}\}$,
the inequality (\ref{eq:prob_bound}) follows trivially.)

 From (\ref{eq:prob_bound}) we determine that
$|l_j/4 - 2^{-j}x|_1<\frac{1}{4}$ holds for all values of $j$ with
probability at least $1 - 4ne^{-k/8}$.

Now consider the following problem:

\noindent\hrulefill
\begin{center}
\begin{tabular}{lp{5in}}
Input: & $l_1,\ldots,l_n\in\{0,1,2,3\}$.\\
Promise: & There exists $x\in\{0,1\}^n$ such that
$\left|l_j/4-2^{-j}x\right|_1 < \frac{1}{4}$ for $j=1,\ldots,n$.\\
Output: & $x$ satisfying the promise.
\end{tabular}
\end{center}
\noindent\hrulefill\vspace{4mm}

The following algorithm solves this problem:
\begin{mylist}{6mm}
\item[1.] Define
\[
A_0 = \left(\!\begin{array}{cc}1&0\\0&1\end{array}\!\right), \rule{4mm}{0mm}
A_1 = \left(\!\begin{array}{cc}1&1\\0&0\end{array}\!\right), \rule{4mm}{0mm}
A_2 = \left(\!\begin{array}{cc}0&1\\1&0\end{array}\!\right), \rule{4mm}{0mm}
A_3 = \left(\!\begin{array}{cc}0&0\\1&1\end{array}\!\right).
\]

\item[2.] Let $x_j =A_{l_j}A_{l_{j-1}}\cdots A_{l_1}[2,1]$ for each $j$, and
output $x = x_n\cdots x_1$.
\end{mylist}

\noindent
Let us now demonstrate that the algorithm is correct.
We note that it is straightforward to show that for a given input
$l_1,\ldots,l_n$ there is at most one $x$ satisfying the promise, and thus
the solution is uniquely determined if the promise holds.
To show that the algorithm computes this $x$ correctly, we prove by induction
on $j$ that $x_j$ is output correctly.
The set $\{A_0,A_1,A_2,A_3\}$ is closed under matrix multiplication, so we
must have that the first column of $A_{l_i}\cdots A_{l_1}$ is either
\[
e_1 := \left(\!\!\begin{array}{c}1\\0\end{array}\!\!\right)\rule{10mm}{0mm}
\mbox{or}\rule{10mm}{0mm}
e_2 := \left(\!\!\begin{array}{c}0\\1\end{array}\!\!\right)
\]
for each $i$.
Thus it suffices to prove that the first column of $A_{l_j}\cdots A_{l_1}$ is
$e_1$ if $x_j = 0$ and is $e_2$ if $x_j = 1$.
The base case is $j = 1$.
Either $x_1 = 0$, in which case the fractional part of $2^{-1}x$ is $0$, or
$x_1 = 1$, in which case the fractional part of $2^{-1}x$ is $1/2$.
By the promise, we must therefore have $l_1 \in\{0,1\}$ in case $x_1 = 0$ and
$l_1 \in\{2,3\}$ in case $x_1 = 1$.
Thus the first column of $A_{l_1}$ is $e_1$ if $x_1 = 0$ and is
$e_2$ if $x_1 = 1$ as required.
Now suppose $x_j,\ldots,x_1$ are output correctly.
We want to show that the first column of
$A_{l_{j+1}}\cdots A_{l_1}$ is $e_1$ if $x_{j+1} = 0$ and is $e_2$ if
$x_{j+1} = 1$.
There are four possibilities for the pair $(x_{j+1},x_j)$ that, along with
the promise, give rise to the following implications:
\[
\begin{array}{l}
x_{j+1}=0,\:x_j=0  \;\Rightarrow\; l_{j+1} \in \{0,1\}\\[1mm]
x_{j+1}=0,\:x_j=1  \;\Rightarrow\; l_{j+1} \in \{1,2\}\\[1mm]
x_{j+1}=1,\:x_j=0  \;\Rightarrow\; l_{j+1} \in \{2,3\}\\[1mm]
x_{j+1}=1,\:x_j=1  \;\Rightarrow\; l_{j+1} \in \{3,0\}\\[1mm]
\end{array}
\]
Suppose $x_j = 0$, implying that the first column of $A_{l_j}\cdots A_{l_1}$
is $e_1$.
If $x_{j+1} = 0$ then either $l_{j+1} = 0$ or $l_{j+1} = 1$, in either case
implying that the first column of $A_{l_{j+1}}\cdots A_{l_1}$ is $e_1$,
as required.
Similarly, if $x_{j+1} = 1$ then either $l_{j+1} = 2$ or $l_{j+1} = 3$, in
either case implying that the first column of $A_{l_{j+1}}\cdots A_{l_1}$ is
$e_2$, as required.
The case $x_j = 1$ is similar.
Thus we have shown that the algorithm operates correctly.

The above algorithm lends itself well to parallelization, following from the
parallel prefix method discussed in subsection~\ref{subsec:copy};
by Theorem~\ref{theorem:pp} all values of
$x_j=A_{l_j}A_{l_{j-1}}\cdots A_{l_1}[2,1]$, $j=1,\ldots,n$ 
can be computed by a single circuit of size $O(n)$ and depth $O(\log n)$ 
(following from the fact that multiplication of the $2\times 2$ matrices, in 
modulo 2 arithmetic, can of course be done by constant-size circuits).

It follows that the entire circuit for approximating the mapping
(\ref{eq:mapping3}) given $k \in O(\log(n/\varepsilon))$ copies of
$\ket{\psi_x}$ has size $O(n\log(n/\varepsilon))$ and depth
$O(\log n + \log\log(n/\varepsilon)) = O(\log n + \log\log(1/\varepsilon))$.
It remains to argue that the circuit operates with error $O(\varepsilon)$.
This follows from standard results based on ideas in \cite{BennettB+97} 
about converting quantum circuits that perform measurements and 
produce classical information with small error probability into unitary 
operations (without measurements) that can operate on data in superposition.
It should be noted that a state $\ket{\psi_x}$ can be conserved throughout 
the computation to ensure that errors corresponding to different values 
of $x$ are orthogonal.


\section{New size bounds for the QFT}
\label{sec:size_bound}

In this section, we prove Theorem~2.
Let $F_{2^n}$ denote the Fourier transform modulo $2^n$, which 
acts on $n$ qubits.
The Hadamard transform is $H = F_2$.

The standard circuit construction for $F_{2^n}$ can be described 
recursively as follows (where the two-qubit controlled-phase shift 
gates of the form $\mbox{c-{\it P\/}}(\theta)$ are defined in 
Section~\ref{sec:notation}).\eee

\noindent\parbox{\textwidth}{
\ni\hspace*{5mm}{\bf Standard recursive circuit description for $F_{2^n}$:}
\begin{enumerate}
\item
Apply $F_{2^{n-1}}$ to the first $n-1$ qubits.
\item
For each $j \in \{1,2,\ldots,n-1\}$, apply $\mbox{c-{\it P\/}}(1/2^{n-j+1})$ 
to the $j\th$ and $n\th$ qubit.
\item
Apply $H$ to the $n\th$ qubit.
\end{enumerate}}
The resulting circuit consists of $n(n-1)/2$ two-qubit 
gates and $n$ one-qubit gates.

Below is a more general recursive circuit description for $F_{2^n}$, 
parameterized by $m\!\in\!\{1,\ldots,n-1\}$.
This coincides with the above circuit when $m = 1$.
When $m > 1$, it can be verified that the circuit does not change 
very much.
It has exactly the same gates, though the relative order of the 
two-qubit gates (which all commute with each other) changes.\eee

\ni\hspace*{5mm}{\bf Generalized recursive circuit description 
for $F_{2^n}$:}
\begin{enumerate}
\item
Apply $F_{2^{n-m}}$ to the first $n-m$ qubits.
\item
For each $j \in \{1,2,\ldots,n-m\}$ and 
$k \in \{1,2,\ldots,m\}$, apply 
$\mbox{c-{\it P\/}}(1/2^{k-j+1})$ to the $j\th$ and $(n-m+k)\th$ qubit.
\item
Apply $F_{2^m}$ to the last $m$ qubits.
\end{enumerate}

Our new quantum circuits are based on this generalized recursive 
construction with $m = \lfloor {n/ 2} \rfloor$, except that 
they use a more efficient method for performing the 
transformation in Step~2.
As is, Step~2 consists of $(n-m)m$ two-qubit gates, which is 
approximately $n^2/4$.
The key observation is that Step~2 computes the mapping which, 
for $x \in \01^{n-m}$ and $y \in \01^m$, takes the state 
$\ket{x}\ket{y}$ to the state 
$(e^{2 \pi i / 2^n})^{x \cdot y}\ket{x}\ket{y}$, 
where $x \cdot y$ denotes the product of $x$ and $y$ 
interpreted as binary integers.  From this, it can be shown that Step~2 can
be computed using  any classical method for integer multiplication in
conjunction with some one-qubit phase shift gates (of the form $P(\theta)$, 
defined in Section~\ref{sec:notation}).

The best asymptotic circuit size for integer multiplication, 
due to Sch\"{o}nhage and Strassen \cite{SchonhageS71}, is 
$O(n \log n \log\log n)$, which can be translated into a reversible 
computation of the same size that we will denote as $S$.
For $x \in \01^{n-m}$ and $y \in \01^m$, $S$ maps the state 
$\ket{x}\ket{y}\ket{0^n}$ to $\ket{x}\ket{y}\ket{x \cdot y}$.
(There are $O(n)$ additional ancilla qubits that are not explicitly indicated.
Each of these begins and ends in state $\ket{0}$.)\eee

\ni\hspace*{5mm}{\bf Improved Step~2 in general circuit description 
for $F_{2^n}$:}
\begin{enumerate}
\item
Apply $S$ to the $2n$ qubits.
\item
For each $k \in \{1,2,\ldots,n\}$ apply $P(1/2^k)$ to the 
$(n+k)\th$ qubit.
\item
Apply $S^{-1}$ to the $2n$ qubits.
\end{enumerate}

Using this improved Step~2 in the generalized recursive circuit 
description for $F_{2^n}$ results in a total number of gates that 
satisfies the recurrence 
\begin{equation} 
T_n = 
T_{\lceil {n/2} \rceil}+ T_{\lfloor {n/2} \rfloor} 
+ O(n \log n \log\log n),
\end{equation}
which implies that $T_n \in O(n (\log n)^2 \log\log n)$.
It is straightforward to also show that the circuit has depth 
$O(n)$ and width $O(n)$ (where ancilla qubits are counted 
as part of the width).


\section{Factoring via logarithmic-depth quantum circuits}
\label{sec:factoring}

In this section we discuss a simple modification of Shor's factoring algorithm
that factors integers in polynomial time using logarithmic-depth quantum
circuits.
It is important to note that we are not claiming the existence of
logarithmic-depth quantum circuits that take as input some integer $n$
and output a non-trivial factor of $n$ with high probability---the method
will require (polynomial time) classical pre-processing and post-processing
that is not known to be parallelizable.
The motivation for this approach is that, under the assumption that quantum
computers can be build, one may reasonably expect that quantum computation
will be expensive while classical computation will be inexpensive.

The main bottleneck of the quantum portion of Shor's factoring algorithm is
the modular exponentiation.
Whether or not modular exponentiation can be parallelized is a long-standing
open question, and we do not address this question here.
Instead, we show that sufficient classical pre-processing allows
parallelization of the part of the quantum circuit associated with the modular
exponentiation.
Combined with logarithmic-depth circuits for quantum Fourier transform, we
obtain the result claimed in Theorem~\ref{theorem:factoring}.

In order to describe our method, let us briefly review Shor's factoring
algorithm, including the reduction from factoring to order-finding.
It is assumed the input is a $n$-bit integer $N$ that is odd and composite.

\begin{mylist}{5mm}
\item[1.] (Classical)
Randomly select $a \in \{2,\ldots,N-1\}$.
If $\op{gcd}(a,N)>1$ then output $\op{gcd}(a,N)$, otherwise continue to
step 2.

\item[2.] (Quantum)
Attempt to find information about the order of $a$ in $\mathbb{Z}_N$:

\begin{mylist}{5mm}
\item[a.] Initialize a $2n$-qubit register and an $n$-qubit register 
to state $\ket{0}\ket{0}$.

\item[b.] Perform a Hadamard transform on each qubit of the first register.

\item[c.] (Modular exponentiation step.)
Perform the unitary mapping:
\[
\ket{x}\ket{0} \mapsto \ket{x}\ket{a^x \bmod N}.
\]

\item[c.]
Perform the quantum Fourier transform on the first register and measure 
(in the computational basis).
Let $y$ denote the result.

\end{mylist}

\item[3.] (Classical)
Use the continued fraction algorithm to find relatively prime integers 
$k$ and $r$ such that $0\leq k<r<N$ and $|y/2^m - k/r|\leq 2^{-2n}$.
If $a^r\equiv 1 \,(\op{mod}\,N)$ then continue to step 4, otherwise
repeat step 2.

\item[4.] (Classical)
If $r$ is even, compute $d=\op{gcd}(a^{r/2}-1,N)$ and output $d$ if it is a
nontrivial factor of $N$. Otherwise go to step 1.

\end{mylist}

The key observation is that much of the work required for the modular 
exponentiation step can be shifted to the classical computation in step 1 
of the procedure.
In step 1, the powers $b_0 = a$, 
$b_1 = a^2 \bmod N$, 
$b_2 = a^{2^2} \bmod N, \ldots$, $b_{2n-1} = a^{2^{2n-1}} \bmod N$ 
can be computed in polynomial-time.
With this information available in step 2, the modular exponentiation 
step reduces to applying a unitary operation that maps 
$\ket{b_0}\ket{b_1} \cdots \ket{b_{2n-1}}\ket{x}\ket{0}$ to 
$\ket{b_0}\ket{b_1} \cdots \ket{b_{2n-1}}
\ket{x}\ket{b_0^{x_0} \cdot b_1^{x_1} \cdots 
b_{2n-1}^{x_{2n-1}} \bmod N}$.
This is essentially an iterated multiplication problem, where one is 
given $2n$ $n$-bit integers \linebreak
$b_0^{x_0}, b_1^{x_1}, \ldots, b_{2n-1}^{x_{2n-1}}$ as input and 
the goal is to compute their product.
The most straightforward way to do this is to perform pairwise 
multiplications following the structure of a binary tree with 
$2n$ leaves.
Each multiplication can be performed with depth $O(\log n)$ and 
size $O(n^2)$.
The underlying binary tree has depth $\log(2n)$ and $2n-1$ internal 
nodes.
Thus, the entire process can be performed with depth $O((\log n)^2)$ 
and size $O(n^3)$.

There are alternative methods for performing iterated multiplication 
achieving various combinations of depth and size.
In particular, it was proved by Beame, Cook and Hoover \cite{BeameC+86}
that a product such as we have above can be computed by $O(\log n)$
depth boolean circuits of size $O(n^5(\log n)^2)$.
While $O(n^5 \log n)$ qubits may seem a high price to pay in order to
save a factor of $O(\log n)$ in the circuit depth, the result has an
interesting consequence regarding simulations of logarithmic-depth quantum
circuits: if logarithmic-depth quantum circuits can be simulated in
polynomial time, then factoring can be done in polynomial time as well.
It should be noted that the circuits of Beame, Cook and Hoover are
not logspace-uniform but rather are polynomial-time uniform; the best known
bound on circuit depth for iterated products in the case of logspace uniform
circuits is $O(\log n\log\log n)$ due to Reif \cite{Reif86}.


\section{Lower bounds}
\label{sec:lowerbounds}

Logarithmic-depth lower bounds for {\em exact\/} computations with two-qubit 
gates are fairly easy to obtain, based on the fact that the state of 
some output qubit (usually) critically depends on every input qubit.
Since, by Eq.~\ref{cemm}, the last qubit of 
$\ket{\psi_{x_{n-1} \ldots\, x_1 x_0}}$ is in state 
${1\over\sqrt{2}}(\ket{0}+e^{2 \pi i (0.x_{n-1} \ldots\, x_1 x_0)}\ket{1})$, 
its value depends on all $n$ input qubits to the QFT when its input 
state is $\ket{x_{n-1} \ldots\, x_1 x_0}$.
The depth of the circuit must be at least $\log n$ for this to be possible.
This lower bound proof applies not only to the QFT, but also to QFS 
computations (which are defined in Section~\ref{sec:notation}).
This is because the output of a QFS on input $\ket{x}\ket{0}$ includes 
the state $\ket{\psi_x}$.

On the other hand, {\em approximate\/} computations can sometimes be 
performed with much lower depth than their exact counterparts.
For example, in Section~\ref{subsection:create}, it is shown that a QFS can be
computed with 
precision $\e$ by a quantum circuit with depth $O(\log\log(n / \e))$.
Note that this is $O(\log\log n)$ whenever $\e \in 1/n^{O(1)}$.
Although this suggests that it is conceivable for a sub-logarithmic-depth 
circuit to approximate the QFT with precision $1/n^{O(1)}$, 
Theorem~\ref{theorem:lowerbounds} implies that this is not possible.
We now prove this theorem.

Let $C$ be a quantum circuit that approximates the {\em inverse\/} QFT 
with precision $1 \over 10$.
In this section, since we will need to consider distances between 
mixed states, we adopt the {\em trace distance\/} as a measure 
of distance (see, e.g., \cite{Fuchs95}).
The trace distance between two states with respective density operators 
$\rho$ and $\sigma$ is given as 
\begin{equation}
D(\rho,\sigma) = \half \mbox{Tr}|\rho - \sigma|,
\end{equation}
where, for an operator $A$, $|A| = \sqrt{A^{\dag}A}$.
For a pair of pure states $\ket{\phi}$ and $\ket{\phi^{\prime}}$, 
their trace distance is $\sqrt{1 - |\bracket{\phi}{\phi^{\prime}}|^2}$, 
which is upper bounded by their Euclidean distance.

On input $\ket{\psi_{x_{n-1} \ldots\, x_1 x_0}}$, the output state of 
$C$ contains an approximation of $\ket{x_{n-1} \ldots\, x_1 x_0}$.
In particular, one of the output qubits of $C$ should be in a 
state that is an approximation of $\ket{x_{n-1}}$ within $1 \over 10$.
Let us refer to this as the {\em high-order\/} output qubit of $C$.
If the depth of $C$ is less than $\log n$ then the high-order output 
qubit of $C$ cannot depend on all $n$ of its input qubits.
Let $k \in \{0,1,\ldots,n-1\}$ be such that the high-order output qubit 
does not depend on the $k^{\mbox{\scriptsize th}}$ input qubit (where 
we index the input qubits right to left starting from 0).
Let $r = n-k-1$.

Set $z = 2^n - 1$, which is $11 \ldots 1 = 1^n$ in binary.
Following Eq.~\ref{cemm2}, $\ket{\psi_z}$ can be written as 
\begin{equation}
\ket{\psi_z} = \ket{\mu_{0.1}} \ket{\mu_{0.11}} \cdots 
\ket{\mu_{0.1^n}}.
\end{equation}
Consider the state $\ket{\psi_{z + 2^r}}$.
Since $z+2^r = 0^{n-r}1^r\,(\op{mod}\,2^n)$, 
\begin{equation}
\ket{\psi_{z + 2^r}} = 
\ket{\mu_{0.1}} \ket{\mu_{0.11}} \cdots \ket{\mu_{0.1^r}} 
\ket{\mu_{0.01^r}} \ket{\mu_{0.001^r}} \cdots \ket{\mu_{0.0^{n-r}1^r}}.
\end{equation}
Note that, on input $\ket{\psi_z}$, the high-order output qubit of $C$ 
approximates $\ket{1}$ with precision $1 \over 10$; whereas, on input 
$\ket{\psi_{z + 2^r}}$, the high-order output qubit of $C$ approximates 
$\ket{0}$ with precision $1 \over 10$.

Now, we consider a state $\ket{\newpsi_z}$, which 
has an interesting relationship with both $\ket{\psi_z}$ and 
$\ket{\psi_{z + 2^r}}$.
Define 
\begin{equation}
\ket{\newpsi_z} = 
\ket{\mu_{0.1}} \ket{\mu_{0.11}} \cdots \ket{\mu_{0.1^r}}
\ket{\mu_{0.01^r}} \ket{\mu_{0.1^{r+2}}} \ket{\mu_{0.1^{r+3}}} 
\cdots \ket{\mu_{0.1^n}}.
\end{equation}
The states $\ket{\newpsi_z}$ and $\ket{\psi_z}$ are 
identical, except in their $k^{\mbox{\scriptsize th}}$ qubit 
positions (which are orthogonal: $\ket{\mu_{0.01^r}}$ vs.\ 
$\ket{\mu_{0.1^{r+1}}}$).
Since the high-order output qubit of $C$ does not depend on its 
$k^{\mbox{\scriptsize th}}$ input qubit, it is the same for 
input $\ket{\newpsi_z}$ as for input $\ket{\psi_z}$.
Therefore, the state of the high-order output qubit of $C$ on input 
$\ket{\newpsi_z}$ is within $1 \over 10$ of $\ket{1}$.

On the other hand, the trace distance between $\ket{\newpsi_z}$ 
and $\ket{\psi_{z + 2^r}}$ can be calculated to be below 0.7712, 
as follows.
The two states are identical in qubit positions $n-1, n-2, \ldots, k$.
In qubit position $k-1$, the two states differ by an angle of 
$\pi \over 4$, in qubit position $k-2$ the two states differ by 
an angle of $\pi \over 8$, and so on.
Therefore, 
\begin{eqnarray*}
\bracket{\newpsi_z}{\psi_{z + 2^r}} & = & 
\bracket{\mu_{0.1^{r+2}}}{\mu_{0.001^r}} 
\bracket{\mu_{0.1^{r+3}}}{\mu_{0.0001^r}}
\cdots
\bracket{\mu_{0.1^n}}{\mu_{0.0^{n-r}1^r}} \\
& = & \cos(\ts{\pi \over 2^2})\cos(\ts{\pi \over 2^3}) 
\cdots \cos(\ts{\pi \over 2^{n-k-1}}) \\
& > & 
\cos(\ts{\pi \over 2^2})\cos(\ts{\pi \over 2^3})\cos(\ts{\pi \over 2^4}) \cdots
\\
& > & 0.6366,
\end{eqnarray*}
where the numerical value for the last inequality is proved in
Lemma~\ref{cosprod} 
(below).
This implies that the trace distance between $\ket{\newpsi_z}$ 
and $\ket{\psi_{z + 2^r}}$ is less than $\sqrt{1 - (0.6366)^2} = 0.7712$.
Since the trace distance is contractive, it follows that the state of the 
high-order output of $C$ on input $\ket{\newpsi_z}$ has trace distance less
than 0.7712 from the state of high-order output of $C$ on input
$\ket{\psi_{z+2^r}}$.
But, by the triangle inequality, this implies that the trace distance between 
$\ket{0}$ and $\ket{1}$ is less than ${1 \over 10} + 0.7712 + {1 \over 10} 
< 1$, which is a contradiction, since $\ket{0}$ and $\ket{1}$ are orthogonal.
This completes the proof of Theorem~\ref{theorem:lowerbounds}.

\begin{lemma}\label{cosprod}
$\cos(\ts{\pi \over 2^2})\cos(\ts{\pi \over 2^3})\cos(\ts{\pi \over 2^4}) 
\cdots > 0.6366$.
\end{lemma}

\loud{Proof:}
We first lower bound the tails of the above infinite product by 
showing that, for any $i \ge 1$, 
$\cos(\ts{\pi \over 2^{i+1}})
\cos(\ts{\pi \over 2^{i+2}})
\cos(\ts{\pi \over 2^{i+3}}) 
\cdots > 1 - \ts{\pi^2 \over 6\cdot4^i}$.
Since, for $t > 0$, $\cos(t) > 1 - {\,\, t^2 \over 2}$, 
\begin{eqnarray*}
\cos(\ts{\pi \over 2^{i+1}})
\cos(\ts{\pi \over 2^{i+2}})
\cos(\ts{\pi \over 2^{i+3}}) 
\cdots & > & 
\left(1 - \ts{\pi^2 \over 2 \cdot 4^{i+1}}\right)
\left(1 - \ts{\pi^2 \over 2 \cdot 4^{i+2}}\right)
\left(1 - \ts{\pi^2 \over 2 \cdot 4^{i+3}}\right) \cdots \\
& \ge & 1 - \half\left(
\ts{\pi^2 \over 4^{i+1}} + 
\ts{\pi^2 \over 4^{i+2}} + 
\ts{\pi^2 \over 4^{i+3}} + \cdots \right) \\
& = & 1 - \ts{\pi^2 \over 6\cdot 4^i}.
\end{eqnarray*}
Now it follows that, for any $i \ge 1$, 
$\cos(\ts{\pi \over 2^2})
\cos(\ts{\pi \over 2^3})
\cos(\ts{\pi \over 2^4}) 
\cdots > 
\cos(\ts{\pi \over 2^2}) \cdots \cos(\ts{\pi \over 2^i})
(1 - \ts{\pi^2 \over 6\cdot 4^i})$. 
Setting $i=8$ in this inequality gives the numerical lower bound.
\qed


\section{Other moduli}
\label{sec:other_moduli}

In this section we discuss the quantum Fourier transform with respect to
moduli that are not powers of 2.
First we briefly sketch a method for performing (in parallel) the QFT for an
arbitrary modulus that uses the QFT with a power of 2 modulus as a black box.
We then discuss Shor's original method for performing the QFT with respect to
a ``smooth'' modulus, and mention how this method may be parallelized as well.

\subsection{Arbitrary moduli}
\label{sec:arbitrary}

Consider the QFT with respect to an arbitrary modulus $m$.
In this subsection we note that it is possible to approximate such a QFT
with high accuracy in parallel using circuits for the quantum Fourier
transform modulo $2^k$ for $k = \lfloor \log m \rfloor + O(1)$.
Using the circuits for the quantum Fourier transform modulo $2^k$ described
previously, we have that for any $\varepsilon$ and modulus $m$ there exists a
depth $O(\log n \log\log (1/\varepsilon))$ quantum circuit that approximates
the QFT modulo $m$ to within accuracy $\e$ for $n = \lceil\log m\rceil$.
The size of the circuit is polynomial in $n + \log (1/\varepsilon)$.

The method exploits a relation between QFTs with different moduli that was
used by Hales and Hallgren \cite{HalesH99} in regard to the Fourier Sampling
problem (see also H{\o}yer \cite{Hoyer00} for an extension and simplified
proof).

The basic components of the technique are as follows: 
\begin{enumerate}
\item
Create a Fourier state with modulus $m$, which is the mapping 
\[
\ket{x}\ket{0} \;\mapsto\; \ket{x}\ket{\psi_x}.
\]
\item
Copy the Fourier state, which is the mapping
\[
\ket{x}\ket{\psi_x}\ket{0} \cdots \ket{0} \;\mapsto\; 
\ket{x}\ket{\psi_x}\ket{\psi_x} \cdots \ket{\psi_x}.
\]
\item
Apply the inverse Fourier transform modulo $2^k$ on each state $\ket{\psi_x}$,
which is the mapping
\[
\ket{x}\ket{\psi_x} \cdots \ket{\psi_x} \;\mapsto\; 
\ket{x}\left(F_{2^k}^{\dagger}\ket{\psi_x}\right)
\cdots\left(F_{2^k}^{\dagger}\ket{\psi_x}\right).
\]
\item
For each (computational basis state) $y$ occurring among the collections
of qubits on which $F^{\dagger}_{k}$ was performed, compute
$\op{round}(y\,m\,2^{-k})$ $\bmod\; m$, and compute the mode of these 
results.
With high probability the result will be $x$.
(A reasonably straightforward calculation shows that observation of
$F_{2^k}^{\dagger}\ket{\psi_x}$ in the computational basis yields some $y$ with
$\op{round}(ym2^{-k}) = x$ with probability greater than $1/2 + \delta$
for some constant $\delta$.)
XOR this result to the qubits in state $\ket{x}$, and reverse the computation
of each $\op{round}(y\,m\,2^{-k})$ and $y$.
With high probability the mapping
\[
\ket{x}\left(F_{2^k}^{\dagger}\ket{\psi_x}\right)
\cdots
\left(F_{2^k}^{\dagger}\ket{\psi_x}\right)
\;\mapsto\;
\ket{0}\left(F_{2^k}^{\dagger}\ket{\psi_x}\right)
\cdots
\left(F_{2^k}^{\dagger}\ket{\psi_x}\right)
\]
has been performed.

\item
Reverse steps 3 and 2, giving the mapping
$\ket{0}\left(F_{2^k}^{\dagger}\ket{\psi_x}\right)
\cdots \left(F_{2^k}^{\dagger}\ket{\psi_x}\right)
\;\mapsto\;\ket{0}\ket{\psi_x}\ket{0}\cdots\ket{0}$.
\end{enumerate}

\noindent
Unfortunately some of the methods used in the power of 2 case (such as
using three-two adders and approximating the individual qubits of the
Fourier basis states) do not seem to work in this case, which results in
the slightly worse depth bound.
The overall size bound increases as well, but is still polynomial.

It is interesting to note that this method does not require the larger
modulus to be a power of 2---effectively the method shows that the QFT
modulo $m$ for any modulus $m$ can be efficiently approximated given
a black box that approximates the QFT modulo $m'$ for any sufficiently large
$m'$.
The technical details regarding this method will appear in the final
version of this paper.


\subsection{Shor's ``mixed-radix'' QFT}
\label{sec:mixed-radix}

We conclude with a brief discussion of Shor's original ``mixed radix''
method for computing the quantum Fourier transform, as it too can be
parallelized (although to our knowledge not as efficiently as the
power-of-2 case discussed previously in this paper).

Shor's original method for computing the QFT is based on the Chinese
Remainder Theorem and its consequences regarding $\mathbb{Z}_m$ for given
modulus $m$.
Here the modulus is $m = m_1 m_2\cdots m_k$ for $m_1,\ldots,m_k$ pairwise
relatively prime and $m_j \in O(\log m)$.
Thus $k\in O(\log m/\log\log m)$ is somewhat less than the number of bits of
$m$, and each $m_j$ has length logarithmic in the length of $m$.
Taking $m_j$ to be the $j\th$ prime results in a sufficiently dense collection
of moduli $m$ for factoring \cite{Shor94} (see Rosser and Schoenfeld
\cite{RosserS62} for explicit bounds and a detailed analysis of such bounds).

Although stated somewhat differently by Shor, the mixed radix QFT method
may be described as follows:

\begin{enumerate}
\item
For $j = 1,\ldots,k$ define $f_j = \frac{m}{m_j}$ and
set $g_j\in\{0,\ldots,m_{j}-1\}$ such that $g_j\equiv f_j^{-1}\;(\bmod\:m_j)$.

\item
Define $C$ to be the (reversible) operator acting as follows for each
$x\in\{0,\ldots,m-1\}$:
\[
C:\ket{x}\mapsto \ket{(x\,\bmod\:m_1),\ldots,(x\,\bmod\:m_k)}
\]

\item
Define $A$ to be a (reversible) operator such that
\[
A:\ket{x_1,\ldots,x_k}\mapsto\ket{g_1 x_1,\ldots,g_k x_k}
\]
for each
$(x_1,\ldots,x_k)\in\{0,\ldots,m_1-1\}\times\cdots\times\{0,\ldots,m_k-1\}$.

\item
Let $F_m$ and $F_{m_j}$ denote the QFT for moduli $m$ and
$m_j$, $j = 1,\ldots,k$, respectively.
Then the following relation holds:
\begin{equation}
\label{eq:mixed-radix}
F_m = C^{\dagger}(F_{m_1}\otimes\cdots\otimes F_{m_k})
AC.
\end{equation}
Thus, to perform the QFT modulo $m$ on $\ket{x}$, first convert $x$
to its {\em modular representation} $(x_1,\ldots,x_k)$ using the
operator $C$, multiply each $x_j$ by $g_j$ (modulo $m_j$), perform the
QFT modulo $m_j$ independently on coefficient $j$ (for each $j$), then apply
the inverse of $C$ to convert back to the ordinary representation of
elements in $\{0,\ldots,m-1\}$.

\end{enumerate}

\ni
The numbers computed in step 1 are used in the standard proof of the
Chinese Remainder Theorem: given $x_1,\ldots,x_k$, we may compute
$x\in\{0,\ldots,m-1\}$ satisfying $x\equiv x_j\,(\bmod\,m_j)$ for each $j$ by
taking $x = \sum_{j=1}^kf_j g_j x_j\,\bmod\,m$.
Thus the operator $C$ can be implemented efficiently, since the mappings
$x\mapsto ((x\,\bmod\,m_1),\ldots,(x\,\bmod\,m_k))$ and
$((x\,\bmod\,m_1),\ldots,(x\,\bmod\,m_k))\mapsto x$ are efficiently
computable (e.g., with size $O(\log^2 m)$ circuits \cite{BachS96}).
In the present case $C$ can be parallelized to logarithmic depth, since each
of the moduli are small.
Similarly, the operator $A$ can be parallelized to logarithmic depth.

To see that the relation (\ref{eq:mixed-radix}) holds, we may simply examine
the action of the operator on the right hand side on computational basis
states:
\begin{eqnarray*}
\lefteqn{C^{\dagger}(F_{m_1}\otimes\cdots\otimes F_{m_k})
AC\ket{x}}\hspace{10mm}\\
& = &
C^{\dagger}(F_{m_1}\otimes\cdots\otimes F_{m_k})
\ket{g_1 x_1,\ldots,g_k x_k}\\
& = &
\frac{1}{\sqrt{m}}C^{\dagger}\sum_{y_1,\ldots,y_k}
\op{exp}(2\pi i g_1 x_1 y_1/m_1)\cdots
\op{exp}(2\pi i g_k x_k y_k/m_k)\ket{y_1,\ldots,y_k}\\
& = & 
\frac{1}{\sqrt{m}}C^{\dagger}\sum_{y_1,\ldots,y_k}
\op{exp}(2\pi i (f_1 g_1 x_1 y_1+\cdots+ f_k g_k x_k y_k)/m)
\ket{y_1,\ldots,y_k}\\
& = &
\frac{1}{\sqrt{m}}\sum_{y_1,\ldots,y_k}
\op{exp}(2\pi i (f_1 g_1 x_1 y_1+\cdots+f_k g_k x_k y_k)/m)
\ket{f_1 g_1 y_1+\cdots+f_k g_k y_k\,(\bmod\,m)}\\
& = &
\frac{1}{\sqrt{m}}\sum_y \op{exp}(2\pi ixy/m)\ket{y}\\
& = & F_m\ket{x}
\end{eqnarray*}

Finally, each of the independent QFTs modulo $m_1,\ldots,m_k$ can of course
be done in parallel.
Here, however, a problem arises if our goal is to parallelize the entire
process.
Originally Shor suggests implementing each of these operations by circuits
of size $m_j$ (not $\log m_j$), since any quantum operation can be computed
by circuits with exponential-size quantum circuits \cite{BarencoB+95}.
This results in a linear-depth circuit overall, although the circuit will
be exact.

However, we may try to compute each $F_{m_j}$ more efficiently.
There are a few possibilities for how to do this, all (apparently)
requiring approximations of each $F_{m_j}$.
First, we may apply the method of Kitaev \cite{Kitaev95} to approximate
these QFTs.
Alternately, we may use the arbitrary modulus method we have proposed in
section~\ref{sec:arbitrary}.
Finally, we have noted that this method works for any two moduli (not just
for the larger modulus a power of 2) so that we may in fact recurse
using the mixed-radix method to approximate each $F_{m_j}$.

In all cases, our analysis has revealed that the mixed radix method
results in worse size and/or depth bounds than the power of 2 method
presented in Section~\ref{sec:depth_bound}.


\section{Conclusion}
\label{sec:conclusion}

We have proved several new bounds on the circuit complexity of 
approximating the quantum Fourier transform, and have applied these bounds to
the problem of factoring using quantum circuits.
There are several related open questions, a few of which we will now
discuss.

First, is it possible to perform the quantum Fourier transform {\em exactly}
using logarithmic- or poly-logarithmic-depth quantum circuits?
The best currently known upper bound on the depth of the exact QFT is linear
in the number of input qubits.

Next, can the efficiency of our techniques be improved significantly?
We have concentrated on asymptotic analyses of our circuits, and we believe
it is certain that our circuits can be optimized significantly for
``interesting'' input sizes (perhaps several hundred to a few thousand qubits).

Finally, the fact that the quantum Fourier transform can be performed in
logarithmic depth suggests the following question: are there interesting
natural problems in BQNC (bounded-error quantum NC) not known to be in NC or
RNC?
For instance, computing the gcd of two $n$-bit integers and computing
$a^b\,\op{mod}\,c$ and $a^{-1}\,\op{mod}\,c$ for $n$-bit integers $a$, $b$,
and $c$ is not known to be possible using polynomial-size circuits with
depth poly-logarithmic in $n$ in the classical setting.
Are there logarithmic- or poly-logarithmic-depth quantum circuits for these
problems?
Greenlaw, Hoover and Ruzzo \cite{GreenlawH+95} list several other problems
not known to be classically parallelizable, all of which are interesting
problems to consider in the quantum setting.


\subsection*{Acknowledgments}

We thank Wayne Eberly for helpful discussions on classical circuit complexity,
and Chris Fuchs and Patrick Hayden for an informative discussion regarding
quantum state distance measures.


\bibliographystyle{plain}

\begin{thebibliography}{10}

\bibitem{AharonovB96}
D.~Aharonov and M.~Ben-Or.
\newblock Polynomial simulations of decohered quantum computers.
\newblock In {\em Proceedings of the 37th Annual Symposium on Foundations of
  Computer Science}, 1996.

\bibitem{BachS96}
E.~Bach and J.~Shallit.
\newblock {\em Algorithmic Number Theory, Volume I: Efficient Algorithms}.
\newblock MIT Press, 1996.

\bibitem{BarencoB+95}
A.~Barenco, C.~H. Bennett, R.~Cleve, D.~DiVincenzo, N.~Margolus, P.~Shor,
  T.~Sleator, J.~Smolin, and H.~Weinfurter.
\newblock Elementary gates for quantum computation.
\newblock {\em Physical Review Letters A}, 52:3457--3467, 1995.

\bibitem{BeameC+86}
P.~Beame, S.~Cook, and H.~J. Hoover.
\newblock Log depth circuits for division and related problems.
\newblock {\em SIAM Journal on Computing}, 15(4):994--1003, 1986.

\bibitem{Bennett73}
C.~H. Bennett.
\newblock Logical reversibility of computation.
\newblock {\em IBM Journal of Research and Development}, 17:525--532, 1973.

\bibitem{BennettB+97}
C.~H. Bennett, E.~Bernstein, G.~Brassard, and U.~Vazirani.
\newblock Strengths and weaknesses of quantum computing.
\newblock {\em SIAM Journal on Computing}, 26(5):1510--1523, 1997.

\bibitem{BonehL95}
D.~Boneh and R.~Lipton.
\newblock Quantum cryptanalysis of hidden linear functions.
\newblock In {\em Advances in Cryptology -- Crypto'95}, volume 963 of {\em
  Lecture Notes in Computer Science}, pages 242--437. Springer-Verlag, 1995.

\bibitem{BrassardH+98}
G.~Brassard, P.~H{\o}yer, and A.~Tapp.
\newblock Quantum counting.
\newblock In {\em Proceedings of the 25th International Colloquium on Automata,
  Languages and Programming}, volume 1443 of {\em Lecture Notes in Computer
  Science}, pages 820--831, 1998.

\bibitem{Cleve94}
R.~Cleve.
\newblock A note on computing quantum {F}ourier transforms by quantum programs.
\newblock Manuscript. Available at {\tt http://www.cpsc.ucalgary.ca/{$\tilde{\
  }$}cleve/papers.html}, 1994.

\bibitem{CleveE+98}
R.~Cleve, A.~Ekert, C.~Macchiavello, and M.~Mosca.
\newblock Quantum algorithms revisited.
\newblock {\em Proceedings of the Royal Society, London}, A454:339--354, 1998.

\bibitem{Cook85}
S.~A. Cook.
\newblock A taxonomy of problems with fast parallel algorithms.
\newblock {\em Information and Control}, 64:2--22, 1985.

\bibitem{Cooley87}
J.~W. Cooley.
\newblock The re-discovery of the fast {F}ourier transform algorithm.
\newblock {\em Mikrochimica Acta}, 3:33--45, 1987.

\bibitem{CooleyT65}
J.~W. Cooley and J.~Tukey.
\newblock An algorithm for the machine calculation of complex {F}ourier series.
\newblock {\em Mathematics of Computation}, 19:297--301, 1965.

\bibitem{Coppersmith94}
D.~Coppersmith.
\newblock An approximate {F}ourier transform useful in quantum factoring.
\newblock Technical Report RC19642, IBM, 1994.

\bibitem{Fuchs95}
C.~Fuchs.
\newblock {\em Distinguishability and Accessible Information in Quantum
  Theory}.
\newblock PhD thesis, University of New Mexico, 1995.
\newblock Los Alamos Preprint Archive, quant-ph/9601020.

\bibitem{GathenG99}
J.~von~zur Gathen and J.~Gerhard.
\newblock {\em Modern Computer Algebra}.
\newblock Cambridge University Press, 1999.

\bibitem{Gauss1866}
C.~F. Gauss.
\newblock Theoria interpolationis methodo nova tractata.
\newblock In {\em Werke III, Nachlass}, pages 265--330. K\"{o}nigliche
  Gesellschaft der Wissenschaften, G\"{o}ttingen, 1866.
\newblock Reprinted by Georg Olms Verlag, Hildesheim, New York, 1973.

\bibitem{Goldreich99}
O.~Goldreich.
\newblock {\em Modern Cryptography, Probabilistic Proofs and Pseudorandomness}.
\newblock Springer, 1999.

\bibitem{GreenlawH+95}
R.~Greenlaw, H.~J. Hoover, and W.~Ruzzo.
\newblock {\em Limits to Parallel Computation}.
\newblock Oxford University Press, 1995.

\bibitem{Grigoriev96}
D.~Grigoriev.
\newblock Testing shift-equivalence of polynomials using quantum machines.
\newblock In {\em Proceedings of the 1996 International Symposium on Symbolic
  and Algebraic Computation}, pages 49--54, 1996.

\bibitem{Grover96}
L.~Grover.
\newblock A fast quantum mechanical algorithm for database search.
\newblock In {\em Proceedings of the Twenty-Eighth Annual ACM Symposium on
  Theory of Computing}, pages 212--219, 1996.

\bibitem{HalesH99}
L.~Hales and S.~Hallgren.
\newblock Quantum {F}ourier sampling simplified.
\newblock In {\em Proceedings of the Thirty-First Annual ACM Symposium on
  Theory of Computing}, pages 330--338, 1999.

\bibitem{HeidemanJ+84}
M.~T. Heideman, D.~H. Johnson, and S.~Burris.
\newblock Gauss and the history of the {F}ast {F}ourier {T}ransform.
\newblock {\em IEEE ASSP Magazine}, pages 14--21, 1984.

\bibitem{Hoyer00}
P.~H{\o}yer.
\newblock {\em Quantum Algorithms}.
\newblock PhD thesis, Odense University, Denmark, 2000.

\bibitem{Kitaev95}
A.~Yu. Kitaev.
\newblock Quantum measurements and the abelian stabilizer problem.
\newblock Manuscript, 1995.
\newblock Los Alamos Preprint Archive, quant-ph/9511026.

\bibitem{LadnerF80}
R.E. Ladner and M.J. Fischer.
\newblock Parallel prefix computation.
\newblock {\em Journal of the ACM}, 27(4):831--838, 1980.

\bibitem{MaslenR95}
D.K. Maslen and D.N. Rockmore.
\newblock Generalized {FFT}s -- a survey of some recent results.
\newblock In L.~Finkelstein and W.~Kantor, editors, {\em Proceedings of the
  DIMACS Workshop on Groups and Computation}, pages 329--369, 1995.

\bibitem{MooreN98}
C.~Moore and M.~Nilsson.
\newblock Parallel quantum computation and quantum codes.
\newblock Los Alamos Preprint Archive, quant-ph/9808027, 1998.

\bibitem{Mosca98}
M.~Mosca.
\newblock Quantum searching and counting by eigenvector analysis.
\newblock In {\em Proceedings of Randomized Algorithms, Workshop of MFCS 98},
  1998.

\bibitem{Ofman63}
Y.~Ofman.
\newblock On the algorithmic complexity of discrete functions.
\newblock {\em Cybernetics and Control Theory}, 7(7):589--591, 1963.

\bibitem{Reif86}
J.~Reif.
\newblock Logarithmic depth circuits for algebraic functions.
\newblock {\em SIAM Journal on Computing}, 15(1):231--242, 1986.

\bibitem{RosserS62}
J.~B. Rosser and L.~Schoenfeld.
\newblock Approximate formulas for some functions of prime numbers.
\newblock {\em Illinois Journal of Mathematics}, 6:64--94, 1962.

\bibitem{SchonhageS71}
A.~Sch\"{o}nhage and V.~Strassen.
\newblock Schnelle multiplikation gro\ss er zahlen.
\newblock {\em Computing}, 7:281--292, 1971.

\bibitem{Shor94}
P.~Shor.
\newblock Algorithms for quantum computation: discrete logarithms and
  factoring.
\newblock In {\em Proceedings of the 35th Annual Symposium on Foundations of
  Computer Science}, pages 124--134, 1994.

\bibitem{Shor97}
P.~Shor.
\newblock Polynomial-time algorithms for prime factorization and discrete
  logarithms on a quantum computer.
\newblock {\em SIAM Journal on Computing}, 26(5):1484--1509, 1997.

\end{thebibliography}

\noindent
(The Los Alamos Preprint Archive may be found at {\tt http://xxx.lanl.gov/}
on the World Wide Web.)


\end{document}